\documentstyle[12pt,fleqn]{article}
\def \be {\begin{equation}}
\def \ee {\end{equation}}
\def\BE{\begin{equation}}
\def\EE{\end{equation}}
\def\BEA{\begin{eqnarray}}
\def\EEA{\end{eqnarray}}
\def\BC{\begin{center}}
\def\EC{\end{center}}

\def\half{\frac{1}{2}}

\def \tr {\hbox{tr}\,}

\def \avrg#1 {\langle #1 \rangle}

\begin{document}

\title{Lyapunov Spectra in SU(2) Lattice Gauge Theory}

\author{
        C. Gong \\
        Physics Department, Duke University \\
        Durham, NC 27708-0305, U.S.A.
	}
\maketitle
\begin{abstract}
We develop a method for calculating the Lyapunov characteristic
exponents of lattice gauge theories. The complete Lyapunov spectrum
of SU(2) gauge theory is
obtained and Kolmogorov-Sinai
entropy is calculated. Rapid convergence with lattice size is found.

\noindent PACS number(s): 11.15.Ha, 11.15.Kc

\end{abstract}


In Ref.1, we have studied chaos in lattice gauge systems by
obtaining the largest Lyapunov exponents. The method used there,
though straightforward, has two drawbacks.
First the results have large error bars because the
exponential divergences of trajectories have fluctuations, which
results in an uncertainty in the determination of the exponential rate
of divergence between trajectories in phase space.
The second drawback of the method is that only the largest Lyapunov
exponent can be obtained, but not the whole Lyapunov spectrum.

There is a well developed method
for calculating Lyapunov spectra of systems with many degrees of freedom,
which is explained in Ref.2 and briefly outlined here.
Given initially a point $q(0)$ in the phase space and $\nu_L$ vectors
$v_{i},i=1,...,\nu_L$ in the
tangent space $T_{q(0)}$, we can integrate the equations of motion
in phase space and simultaneously
the evolution equations for small perturbations in tangent space to
obtain $q(t)$ and $v_{i}(t) \in T_{q(t)}$. At regular time intervals $k\tau$,
the Gram-Schmidt orthonormalization is applied to the tangent
vectors $v_{i}$. The
scaling factors $s_{i}$ obtained by this procedure determine the
Lyapunov exponents as follows,
\be
\lambda_{i}=\lim_{n\rightarrow \infty}
        \sum_{k=1}^{n}\frac{\ln s_{i}^{k}}{\tau},
\ee
where $n$ is the number of iterations performed.
The time needed to obtain the largest
Lyapunov exponent
depends on how fast the exponents converge with increasing $n$.

One necessary condition in the above procedure is that
we are able to identify tangent spaces at different points of phase space.
When the phase space is Euclidean,
the natural identification is used and the problem is trivial.
For lattice gauge theories, where the phase space is not Euclidean,
we must confront this problem explicitly.
In this letter two different approaches are proposed to
obtain the Lyapunov spectrum in lattice gauge theories.
In the first approach, we work in phase space and
avoid the problem by embedding the curved  phase space
into a larger dimensional Euclidean space. This approach
is most useful for SU(2) gauge theory. The method is first tested
on a $10^{3}$ lattice to
calculate the two largest Lyapunov exponents and then applied to obtain
the complete Lyapunov spectrum on lattices of size $1^{3}$,
$2^{3}$ and $3^{3}$, where the scaling
behavior is observed. A second more general approach
is to construct the tangent space. The results
of this method for SU(2)
are not as satisfactory as the first one. The advantage of this second method,
however, is that it can be directly applied to other gauge theories.


The SU(2) lattice
gauge theory is defined by the following Hamiltonian \cite{kogut,chin},
\begin{equation}
H=\sum_{l}\frac{1}{2} E^{a}_{l} E^{a}_{l}+4
 \sum_{p}(1-\half \tr U_{p}),
\end{equation}
where $E_{l}$ are electric fields and $U_{p}$ are plaquette variables
which are ordered products of 4
link variables $U_{l}$. The lattice spacing as well as the coupling
constant have been scaled to unity: $a=g=1$, and the
only parameter of the system is the scaled
average energy per plaquette $g^{2}Ea$, which we will refer as $E$ in the
following.
The temporal axial gauge adopted here does not fix the
gauge completely
and the system is invariant under arbitrary
time-independent gauge transformations.
The total phase space is the direct product of phase space for each link.
The latter itself is the direct product $R^{3}$xSU(2), where
the two subspace are for $E_{l}$ and
$U_{l}$, respectively.
The SU(2) group manifold is isomorphic to
a three dimensional sphere, which can easily be embedded in a
four dimensional Euclidean
space. This is easily achieved if we use the quaternion representation,
where a link variable $U_{l}$ is represented
by a 4-vector $ (u_{0 l},{\bf u_{l}})$
in Euclidean space with the condition of unit norm,
i.e., $u_0^2+{\bf u}^2=1$.
Then, for each link,
phase space forms a six dimensional subspace of a  seven dimensional
Euclidean space.
One way to implement the rescaling method
is to study the evolution of vectors in the corresponding
seven dimensional tangent space.
In order to incorporate the condition
of unit norm of $U_{l}$,
the initial conditions must be carefully chosen.
In the following, we choose a slightly different
way to implement the method.

Instead of following the evolution of $\nu_L$ vectors in the tangent space,
we study $\nu_L+1$ trajectories $z_{i},i=0,...\nu_L$ in phase space.
The trajectory $z_0(t)$ is called the reference trajectory.
At regular time intervals $t=k\tau$,
$\nu_L$ vectors are formed from these trajectories by
\BE
d_i = z_i(t)-z_0(t), i=1,...,\nu_L.
\EE
These vectors are treated as vectors in Euclidean space.
The normal Gram-Schmidt
method is then used to obtain the Lyapunov numbers $\lambda_{i}$ as in (1).
A small difference here is that we do not scale the norm of the
vectors $D_i \equiv  |d_i|$ to unity,
but to a chosen small value $D_{0}$. The scale factor is referred as $s_{i}$.
After this procedure, all the trajectories except the reference trajectory
assume their new positions in phase space.
Although time evolution itself conserves the norm of $U_{l}$, Gauss' law,
as well as
the gauge condition,
the Gram-Schmidt rescaling procedure violates them slightly,
because they form non-linear constraints.
The condition of unit norm is easily imposed by hand. The violation of
Gauss' law is expected not to be serious, the
reason of which is as follows.
If the values $D_i$ are small, then the violation
of Gauss' law is of second order in $D_i$. If we limit ourselves
to sufficiently small $D_i$, then the violation of Gauss' law in each
rescaling step is negligible.
Remembering also that the evolution of the system respects Gauss' law,
the violation does not increase with time. On the other hand, the
next rescaling decreases the previous violation of Gauss' law
by a scale factor $s_i$, so the violations do not accumulate. The
same argument applies to small changes in the choice of gauge.

To test our method,
we have applied it to the SU(2) theory and measured
the two largest Lyapunov exponents.
We indeed find that the violation of Gauss' law remains
of the order of $10^{-6}$, which gives us confidence
in the method. The result of a typical run for the
Lyapunov exponents is shown in Fig.1
for a configuration with scaled energy $E=4.06$.
The solid line corresponds to the largest exponent and the dotted line to
the second largest one. They converge at $t \approx 100$. From
our previous
study \cite{muller} we know that the time scale
for saturation of the distance $D(t)$
between two configurations in the case without rescaling is
about 30 at the same energy. The result obtained with our new, improved
method, $\lambda_{1}=0.667$, is very close to,
but slightly lower than, our previous result $E/6=4.06/6=0.68$.
We note that the result
for the Lyapunov exponent generally converges from above, i.e. the
Lyapunov exponents are overestimated when the trajectories are not
followed for sufficiently long times.  We also observe
that $\lambda_{2}$ is almost identical to $\lambda_{1}$.
The reason is that, as we are going to show next, there exists a whole
Lyapunov spectrum which forms a continuous curve in the large
volume limit.


This method in principle can be used to obtain the Lyapunov spectrum
of a SU(2) lattice gauge system of an arbitrary size.
Practically, the computing requirements limit us to rather small lattices.
We have numerically studied lattices of size $N^{3}$, with $N=1,2,3$.
Fortunately, as we will show,
the spectrum starts to scale as early as at size $N=3$,
which permits us to extrapolate the results to the thermodynaimc limit
without actually going to a larger lattice.

We have obtained the complete
Lyapunov spectrum for systems on $1^{3}$ and $2^{3}$ lattices.
On a $N^{3}$ lattice the phase space dimension is $3^{2}N^{3}\times 2
=18N^{3}$,
because there are three space directions
and three color directions at each
site for magnetic and electric fields.
Hence on a $2^3$ lattice there
are 144 Lyapunov exponents which are shown in Fig.2.
Measurements are performed at different times and convergence
in time is observed.
In Fig.2, two measurements at $t=200$ ( the crosses)
and  $t=1000$ (the solid triangles) are shown.
The later has a smaller fluctuaction.
We can see that the spectrum is divided into three equal parts.
The first one third of Lyapunov exponents
are positive, while the second one third are all zero
(they are not exactly vanishing at $t=1000$,
but they are clearly seen to converge to zero). The last
one third of exponents are approximately
the negative of the first one third.
The vanishing Lyapunov exponents correspond to the
conservation of charge (Gauss' law) and the
gauge degrees of freedom at each lattice site.
Thus our results confirm the general properties of Lyapunov spectra
\cite{benettin}.
Our results also implies that besides the total energy and static
color charge at each lattice site,
there is no other conservative quantity in the system.
The results for a $1^3$ lattice are basically the same, but the Lyapunov
spectrum consists only of 18 numbers.
For a run with energy $E=2.632$,
 the value of the largest Lyapunov
exponent is 0.44.
We can compare this result with the result obtained earlier for the model
of spatially constant Yang-Mills potentials \cite{chirikov},
where it is found $\lambda_{1} =hE^{1/4}$,
with $h \approx 0.38$. Inserting $E=2.632$ we get
$\lambda_{1}=0.48$. Taking into account the uncertainties in $h$,
we think the results obtained in these two entirely different
approaches are surprisingly close to each other.

In Fig.3 we show the scaling of the Lyapunov spectrum,
where we compare the results from lattices of different sizes.
The solid line is for a $3^{3}$ lattice, the dotted line
is for a $2^{3}$ lattice
and the solid squares are for a $1^{3}$ lattice.
In the $3^{3}$ case we only calculated
the positive Lyapunov exponents.
In each case, the
initial configurations were chosen as entirely random magnetic
fields and zero initial electric
fields.
In order to observe the scaling of Lyapunov spectrum with respect
to lattice size, the Lyapunov exponents are
scaled with respect to the largest exponent $\lambda_1$.
The indices for the Lyapunov numbers are scaled to the total
number of Lyapunov numbers, i.e. 18 for a $1^{3}$ lattice,
144 for a $2^{3}$ lattice and 486
for a $3^{3}$ lattice.
The solid squares do not yet scale very well with the lines,
but results for $N=2$ and $N=3$ coincide nicely,
exhibiting an early scaling behavior.
The scaled Kolmogorov-Sinai entropy, i.e. the sum
over all positive Lyapunov exponents, is:
\be
\alpha = \frac{\sum_{i}\lambda_{i}}{N^{3}\lambda_{1}} \approx 2.0,
\qquad {\rm for }\qquad N=2,3.
\ee
The reason of this early scaling behavior has not been fully understood yet.

Now we study
the energy dependence of the Lyapunov spectrum.
For initial condition, we chose the link variables
$U_{l}=\cos(\rho/2)-i{\bf n}\cdot{\bf \tau} \sin(\rho/2)$
randomly and
the total energy is varied  by selecting $\delta$ which
In the top part of Fig.4, the Lyapunov spectra corresponding to
three different energies on a $3^3$ lattice are plotted in the scaled form.
The solid triangles correspond to an average plaquette
energy of $E=4.25$, the squares
correspond to $E=3.21$ and the crosses correspond to $E=1.67$.
We see that at this small lattice size, the spectrum does not
yet scale with energy, or in another words, the scaled summation
$\alpha$ is a function of $E$.
In the above three cases, $\alpha = 2.0, 1.7, 1.3$, respectively.
The energy dependence of $\alpha$ on energy is shown in the lower
part of Fig.4, where we see $\alpha$ increases with energy.
In the previous study \cite{muller}, we found that if the lattice is
large enough ($N \ge 6$),
the largest Lyapunov exponent $\lambda_{1}$ depends linearly on
energy, $\lambda_{1}a \approx \frac{1}{6}g^{2}Ea$,
i.e., $\lambda_{1}$ does not depend on the lattice cutoff $a$.
(Here and in the remaining
of this paragraph, we keep $g$ and $a$ explicitly
in order to make connections with
physics in the continuum limit.)
If we assume that
in the large lattice limit, the other Lyapunov exponents are
also independent of the lattice cutoff $a$,
from dimensional consideration,
they can only depend linearly on $g^{2}E$.
Then in the large volume limit, we expect that
$\alpha$ is independent
on energy and is a universal number, which is approximately 2.
We obtain the Kolmogorov-Sinai
entropy density in the thermodynamic limit
\be
{\dot\sigma}= {1\over (Na)^3} \sum_i \lambda_i
=\alpha \lambda_{1} \approx \frac{1}{9}g^{2}\varepsilon,
\ee
where $\varepsilon=3E/a^{3}$ is the energy density.
To represent the entropy growth rate in terms of temperature,
we make use of the thermodynamic relation
$\sigma T = \varepsilon + P$. Using the fact that the partition
function $Z(T,g,a)$ in the classical limit of the lattice gauge
theory depends only on the
combination of $g^{2}Ta$, it is easy to prove $\varepsilon=3P$ \cite{note1}.
Thus we have $\sigma T  = {4\over 3}\varepsilon$,
and we find that the characteristic entropy growth rate for SU(2) is
given by
\be
{{\dot\sigma}\over\sigma}
\approx {1\over 9} g^2 {\varepsilon\over\sigma}
= {1\over 12} g^2 T.
\ee
Its inverse gives an estimation of the thermalization time for
highly excited SU(2) gauge fields.

The above method is quite successful for SU(2) gauge theory, but
it is not obvious how to apply it to SU(3) gauge fields. The reason is
that this method relies on the quaternion representation, which is
quite special for SU(2) group.
In this final part, we propose
a more general method which
can be used to study other gauge theories such as
SU(3).

We try to construct a tangent
space upon the curved phase space \{$E_{l},U_{l}$\} and then study
the evolution of vectors in this space.
The tangent space of $E_{l}$ is simple,
in which a vector is just $\delta E_{l}$.
We shall be careful about how to specify a vector in the
tangent space on $U_{l}$. Here, to be consistent with our
definition of the congugate momenta, $E_{l}$,
as the left group generators \cite{kogut,chin},
we define a vector $b_l$ in the tangent space of $U_l$ as
\BE
\delta U_{l}=i b_{l} U_{l}.
\EE
A vector in the
complete tangent space is the direct product of $b_{l}$ and
$\delta E_{l}$. The linear evolution equations for $b_{l}$ and $\delta E_{l}$
are derived from the equations of motion of $U$ and $E$,
\BEA
\frac{d}{dt}b_{l}&=&\delta E_{l}+i [E_{l},b_{l}], \nonumber \\
\frac{d}{dt} \delta E_{l}^{a} &=&
	\sum_{p(l),m,b} (-\half) \tr U_{p,l,m}^{a,b} b_{m}^{b},
\EEA
where $ U_{p,l,m}^{a,b}$ is obtained from
$U_{p}$, which contains link $l$ and $m$ in one of its four
four positions, by substituting $U_{l}$ by $\tau^{a}U_{l}$ and then $U_{m}$
by $\tau^{b} U_{m}$, where $\tau^{a}$ are Pauli matrices.
These equations can be integrated along with
the equations of motion of $E_{l}$ and $U_{l}$ in the
phase space.

We have tried this method on SU(2) gauge theory, where the results can be
compared with the results obtained
by our first method. Good agreement are found
for large positive exponents;
but for the smaller ones, the convergence of the new method
is not satisfactory.
Future work is still required here.

In conclusion, we have developed a method to study the Lyapunov
characteristic numbers of lattice gauge theory. The whole Lyapunov spectrum
of SU(2) is
obtained explicitely and the scaling behavior is observed.
The Kolmogorov-Sinai
entropy is obtained.

$Acknowledgements$:
The author likes to thank B. M\"uller and S. G. Matinyan
for very useful discussions during this work.
This work was supported in part
by the U.S. Department of Energy (Grant No. DE-FG05-90ER40592) and
by a computing grant from the North Carolina Supercomputing Center.

\newpage
\bigskip
{\noindent \Large \bf Figure Captions}
\bigskip

\begin{itemize}

\item[Fig.1]
The two largest Lyapunov exponents for SU(2) determined by the rescaling
method. The average of the logarithmic scaling factors $s_i^{(k)}$
approaches the limit from above.

\item[Fig.2]
Complete spectrum of 144 Lyapunov exponents for SU(2) gauge theory
on a $2^3$ lattice. The trajectories were followed up to time $t/a=200$
(crosses) and $t/a=1000$ (triangles). The central third fraction of
Lyapunov exponents (enclosed between the vertical dashed lines)
corresponds to the unphysical degrees of freedom that describe gauge
transformations and deviations from Gauss' law. These exponents
converge to zero in the limit $t\to\infty$.

\item[Fig.3]
Scaling of the Lyapunov spectrum with lattice size $N$. The solid line
corresponds to a $3^3$ lattice; the dashed line is for a $2^3$ lattice.
Only the positive Lyapunov exponents are shown.
The exponents $\lambda_i$ are scaled with the maximal Lyapunov exponent
$\lambda_{max}=\lambda_{1}$
for each lattice size, and the index $i$ is scaled with total
number of Lyapunov number $18N^3$.

\item[Fig.4]
Energy scaling on a $3^3$ lattice.
In the top part, the positive part of the
spectra for three different energies are shown.
The solid triangles corresponds to an average plaquette
energy of 4.25, the squares
corresponds to 3.21 and the crosses corresponds to 1.67.
They do not scale with energies.
In the lower part, the scaled Kolmogorov-Sinai entropy $\alpha$
is shown as a function of energy. In the large $N$ limit, we expect
it to be a horizontal line.

\end{itemize}


\begin{thebibliography}{99999}
\bibitem{note1} Here we take $g$ as an parameter which is
independent on temperature $T$
and lattice spacing $a$. It is possible and probably
more desirable to think $g$ as
an effective coupling which runs with both $T$ and $a$.
In this latter case, the
simple energy-pressure relation $\varepsilon =3P$ no longer holds. But
the change shall not be very large, so our estimate of K-S entropy
density (6) still holds to a good approximate.
\end{thebibliography}
\end{document}